\documentclass[%
 reprint,
 aps,
]{revtex4-1}

\usepackage{graphicx}
\usepackage{dcolumn}
\usepackage{bm}
\usepackage{hyperref}

\usepackage{comment}
\usepackage{amsmath, amssymb, mathrsfs, slashed}
\usepackage{multirow, IEEEtrantools}

\usepackage{tabularx}
\usepackage{xcolor}

\usepackage{acro}

\DeclareAcronym{tov}{
  short=TOV,
  long=Tolman-Openheimer-Volkoff,
}
\DeclareAcronym{sm}{
  short=SM,
  long=standard model,
}
\DeclareAcronym{ns}{
  short=NS,
  long=neutron star,
}
\DeclareAcronym{bns}{
  short=BNS,
  long=binary neutron star,
}
\DeclareAcronym{pns}{
  short=PNS,
  long=proto-neutron stars,
}
\DeclareAcronym{hs}{
  short=HS,
  long=hybrid star,
}
\DeclareAcronym{bh}{
  short=BH,
  long=black hole,
}
\DeclareAcronym{qcd}{
  short=QCD,
  long=quantum chromodynamics,
}
\DeclareAcronym{pqcd}{
  short=pQCD,
  long=perturbation quantum chromodynamics,
}
\DeclareAcronym{lqcd}{
  short=lQCD,
  long=lattice quantum chromodynamics,
}
\DeclareAcronym{eos}{
  short=EOS,
  long=equation of state,
}
\DeclareAcronym{nsm}{
  short=NSM,
  long=neutron star matter,
}
\DeclareAcronym{nm}{
  short=NM,
  long=nuclear matter,
}
\DeclareAcronym{ddb}{
  short=DDB,
  long=density depended couplings with Bayesian analysis,
}
\DeclareAcronym{rmf}{
  short=RMF,
  long=relativistic mean field,
}
\DeclareAcronym{nro}{
  short=NRO,
  long=non-radial oscillation,
}
\DeclareAcronym{ai}{
  short=AI,
  long=artificial intelligence,
}
\DeclareAcronym{gw}{
  short=GW,
  long=gravitational wave,
}
\DeclareAcronym{gr}{
  short=GR,
  long=general relativity,
}
\DeclareAcronym{nicer}{
  short=NICER,
  long=Neutron Star Interior Composition ExploreR,
}
\DeclareAcronym{hp}{
  short=HP,
  long=hadronic phase,
}
\DeclareAcronym{mp}{
  short=MP,
  long=mixed phase,
}
\DeclareAcronym{qp}{
  short=QP,
  long=quark phase,
}
\DeclareAcronym{njl}{
  short=NJL,
  long=Nambu--Jona-Lasinio,
}
\DeclareAcronym{ml}{
  short=ML,
  long=machine learning,
}
\DeclareAcronym{nl}{
  short=NL,
  long=non linear,
}
\DeclareAcronym{pca}{
  short=PCA,
  long=principal component analysis,
}
\DeclareAcronym{qnm}{
  short=QNM,
  long=quasi-normal mode,
}
\DeclareAcronym{hqpt}{
  short=HQPT,
  long=hadron-quark phase transition,
}
\DeclareAcronym{loff}{
  short=LOFF,
  long=Larkin-Ovchinnikov-Fulde-Ferrell,
}
\DeclareAcronym{csc}{
  short=CSC,
  long=color-superconductivity,
}
\DeclareAcronym{cfl}{
  short=CFL,
  long=color-flavor locked,
}
\DeclareAcronym{ur}{
  short=UR,
  long=universal relation,
}
\DeclareAcronym{npe}{
  short={\it npe},
  long=neutron-proton-electron,
}
\DeclareAcronym{rhic}{
  short=RHIC,
  long=relativistic heavy-ion collider,
}
\DeclareAcronym{lhc}{
  short=LHC,
  long=large hadron collider,
}
\DeclareAcronym{eft}{
  short=EFT,
  long=effective field theory,
}
\DeclareAcronym{pnm}{
  short=PNM,
  long=pure-neutron matter,
}
\DeclareAcronym{ceft}{
  short=CFT,
  long=chiral effective field theory,
}
\DeclareAcronym{ci}{
  short=CI,
  long=confidence interval,
}
\DeclareAcronym{fair}{
  short=FAIR,
  long=Facility for Antiproton and Ion Research,
}
\DeclareAcronym{frib}{
  short=FRIB,
  long=Facility for Rare Isotope Beams,
}

\begin{document}

\title{Constraining the neutron star equation of state by including the isoscalar-vector and isovector-vector coupling using the Bayesian analysis}

\author{Deepak Kumar}%
\email{deepak.kumar@iopb.res.in}
\author{Pradip Kumar Sahu}%
\email{pradip@iopb.res.in}
\affiliation{%
Institute of Physics, Sachivalaya Marg, Sainik School P.O., Bhubaneswar 751 005, India \\
Homi Bhabha National Institute, Anushakti Nagar, Mumbai, 400094, India
}%

\date{\today}

\begin{abstract}
We constrain the nuclear matter equation of state within the relativistic mean field model by including the isoscalar-vector and isovector-vector coupling term at a fundamental level using the Bayesian analysis. We used the nuclear saturation properties and recent astrophysical observations to constrain the dense matter equation of state. We obtained about 20000 sets of equations of states out of sampling about 60 millions sets of equations of states. All 20000 equations of states satisfy nuclear matter saturation properties at saturation densities and produces high mass neutron stars. In our findings, we find that the non-zero value of isoscalar-vector and isovector-vector coupling parameter and negative value of sigma meson self-coupling stiffen the equation of state. Our sets of equations of state produces neutron stars of mass larger than 2.5 M$_{\odot}$ to include the recent gravitational waves observation GW190419.
\end{abstract}

\keywords{Strong interacting matter, Equation of state, Neutron star}

\maketitle

\section{Introduction} \label{sec:introduction}
The typical baryon density at the core of \ac{ns}s is about a few times the nuclear saturation density ($\rho_0 \sim 0.16\ {\rm fm}^{-3}$) \cite{Rhoades:1974fn}. However, at such high densities, the many-body interactions are poorly known, which gives considerable uncertainty in the theoretical description of the high-density behavior of dense matter \ac{eos}. The theoretical understanding of the \ac{eos} of dense nuclear matter has been one of the main frontiers in nuclear physics in recent decades. \ac{ns}s are the natural astrophysical laboratories to study strong interactions under extreme conditions, which are not possible on Earth, such as extreme densities, extreme pressure, etc. Recent advancements in multi-messenger astronomy, combining observations from \ac{gw} \cite{LIGOScientific:2017vwq, LIGOScientific:2018cki, LIGOScientific:2020aai, LIGOScientific:2020zkf}, X-ray \cite{Miller:2019cac, Miller:2021qha, Riley:2019yda}, and gamma-ray astronomy \cite{LIGOScientific:2017zic}, along with nuclear experiments such as ASY-EoS experiment at GSI \cite{Russotto:2016ucm} and heavy-ion collisions (HIC) \cite{Senger:2021tlg, Zhang:2020azr, Sahu:2000us} have significantly improved our ability to probe these strong interactions \cite{Holt:2014hma, Lattimer:2012nd}. In this context, a comprehensive study of symmetry energy and nuclear saturation properties helps in bridging the gap between these experiments and astrophysical observations, and it enhances our understanding of fundamental nuclear interactions at such extreme densities \cite{Li:2008gp}. To accurately determine these properties, a combination of multi-messenger astrophysics, nuclear experiments, and advanced statistical methods is essential. 

To date, several mass-radius measurements have been done from the X-rays observations \cite{Watts:2016uzu, Ozel:2016oaf}, \ac{nicer} observations of pulsars e.g. PSR J0740+6620 \cite{Miller:2021qha, Dittmann:2024mbo}, and PSR J0030+0451 \cite{Miller:2019cac, Riley:2019yda, Vinciguerra:2023qxq} and the \ac{gw} observation, GW170817, \cite{LIGOScientific:2017vwq, LIGOScientific:2018cki, De:2018uhw} which help to determine the characteristics of strong interactions at such high densities relevant for NS and also for finite nuclei at extreme isospin asymmetry. The recent binary neutron star merger event GW190814 \cite{LIGOScientific:2020zkf, LIGOScientific:2020aai} gives challenges to revisit the theoretical models for high-mass NSs \cite{Dey:2024vsw, Du:2025zjk, Wu:2025uck, Pattnaik:2025zet, Zhao:2020dvu, Folias:2024upz} and/or low-mass black holes \cite{Green:2020jor}. To overcome promptly to such complex problem, one can adopt the powerful Bayesian analysis, which allows the systematic incorporation of experimental data \cite{Xie:2019sqb, Malik:2022zol}, astrophysical observations \cite{LIGOScientific:2017vwq, LIGOScientific:2018cki, LIGOScientific:2020aai, LIGOScientific:2020zkf, Miller:2019cac, Miller:2021qha, Riley:2019yda}, and theoretical models \cite{Wu:2011zzb, Todd-Rutel:2005yzo, Dutra:2014qga, Lalazissis:2009zz, Horowitz:2000xj, Sahu:2000ut, Sahu:1993db, Burgio:2001mk, Burgio:2002sn} to constrain the nuclear \ac{eos}s. This approach has been particularly effective in analysing GW events (e.g., GW170817), NS mass-radius trajectories (\ac{nicer}), and heavy-ion collision data, leading to improved constraints on nuclear \ac{eos}. Furthermore, the \ac{eos} that satisfies these astrophysical observations obtained from the Bayesian approach must be consistent with the nuclear saturation properties, e.g. symmetry energy \cite{Lattimer:2014scr} and symmetric matter \cite{Leonhardt:2019fua} at saturation density. The symmetry energy, which is a measure of the difference between the pure neutron matter energy and symmetric nuclear matter energy at a constant baryon density, is tightly constrained by a concordance achieved from nuclear experiments \cite{Garg:2018uam, Shlomo:2006ole, Oertel:2016bki, Li:2020ass}, and directly influences NS radii, tidal deformability, non-radial oscillation modes, and love number for a binary system \cite{Kumar:2024fui, Kumar:2023rut, LIGOScientific:2017vwq, LIGOScientific:2017zic, Patra:2020wjy}. Similar to symmetry energy, the symmetric nuclear matter serves as the fundamental benchmark to model the \ac{eos} of dense nuclear matter as it influences the bulk properties of nuclear matter at typical nuclear densities, including key parameters such as saturation density, compressibility, and binding energy per nucleon \cite{sym16020215} and oscillation modes of \ac{ns} \cite{Kumar:2024fui, Patra:2025xtd, Sahu:2001iv, Sahu:1993ej}.

At present, the Walecka-type models, where baryons interact via exchange of mesons, are the prominent ones near the nuclear saturation density, which provide the \ac{rmf} theory framework for describing dense nuclear matter \cite{Walecka:1974qa, Serot:1997xg, Serot:1979cc, Ring:1996qi}. Various signatures of nuclear matter, such as attractive interactions, repulsive interactions, are described through the exchange of various mesons \cite{Walecka:1974qa, Serot:1997xg, Serot:1979cc, Ring:1996qi}. We discuss more about these effects in later sections. In recent years, several studies have been done with such kinds of models with considering various interaction terms at the Lagrangian level \cite{Wu:2011zzb, Todd-Rutel:2005yzo, Dutra:2014qga, Lalazissis:2009zz, Horowitz:2000xj}. The interactions between baryons take place through Yukawa interaction terms \cite{Peskin:1995ev, Yukawa:1935xg}. Furthermore, the isoscalar meson self-interactions are required for the appropriate symmetric nuclear matter while other additional terms e.g. isovector-vector and isoscalar-isovector mesons interactions are necessary to modify the density dependence of the symmetry energy and the neutron skin thicknesses of heavy nuclei as well as to modify the stiffness of the \ac{eos} while keeping the nuclear saturation properties ineffective \cite{Horowitz:2000xj, Wu:2011zzb, Todd-Rutel:2005yzo}. We consider here the same interaction in the Lagrangian and give a set of \ac{eos}s which satisfy the nuclear saturation properties and astrophysical observations. On the other hand, several other models also have been used to explain the finite nuclei and astrophysical observations where the interaction couplings are density dependent \cite{Typel:1999yq, Typel:2005ba, Lalazissis:2005de, Typel:2009sy, Malik:2022zol} and with a large number of interactions terms at the Lagrangian level  \cite{Kumar:2017wqp, Singh:2014zza, Furnstahl:1996wv} and are not our current interest. 

In the present study, we consider the \ac{rmf} \cite{Behera:2007uqk} model of FSUGold interaction terms in the Lagrangian density \cite{Todd-Rutel:2005yzo, Horowitz:2000xj, Wu:2011zzb} to describe accurately the saturation properties of nuclear matter and finite nuclei \cite{Sahu:2009um} as well as we take a Lorentz covariant extrapolations for the \ac{eos} of dense matter at few times nuclear saturation density to accommodate high mass \ac{ns}s \cite{LIGOScientific:2017vwq, LIGOScientific:2018cki, LIGOScientific:2020aai, LIGOScientific:2020zkf, Miller:2019cac, Miller:2021qha, Riley:2019yda}. Furthermore, the important aspect of this work is to study the coupling strength ($\Lambda_{\omega\rho}$) between isovector-vector meson ($\vec{\rho}_{\mu}$) and isoscalar-isovector meson ($\omega_{\mu}$) \cite{Dutra:2014qga, Todd-Rutel:2005yzo, Lalazissis:2009zz} using the Bayesian approach. As an exchange of isoscalar-vector mesons, ($\omega_{\mu}$ gives the repulsive interaction and hence the stiff \ac{eos}, we find that the \ac{eos} gets stiffer for the non-zero coupling parameter $\Lambda_{\omega\rho}$. As the value of this parameter is large, the \ac{eos} becomes stiffer. In this work, we discuss this particular parameter in such a way that the \ac{eos} set must satisfy nuclear saturation properties as well as recent astrophysical observations.

This work is organized as follows. In Sec. \ref{sec:formalism}, a detailed overview of a \ac{eos} within the hadrodynamic model, the \ac{rmf} model at the zero temperature limit is given. In Sec. \ref{sec:bayesian_analysis}, we give a brief description of nuclear saturation properties evaluated for the present model and the description of Bayesian analysis. In Sec. \ref{sec:results_and_discussions}, we discuss our findings of the present study. In the end, we give the summary and conclusion in section \ref{sec:summary_and_conclusion}. 

\section{Formalism} \label{sec:formalism}
Here we are focusing on the neutron star matter, i.e. nuclear matter at high densities governed in the Walecka-type models where the hadrons are moving in the mesonic soup. They are interacting through an exchange of various mesons. The exchange of a scalar meson gives rise to the attractive interaction, while an exchange of a vector meson gives rise to the repulsive interaction. We are also considering the isovector-vector meson, which defines the breaking of isospin symmetry in the system. The Lagrangian, which we are considering here, is given as follows \cite{Mishra:2001py, Tolos:2017lgv}
\begin{IEEEeqnarray}{rCl}
{\cal L} = {\cal L}_{\rm B}^{\rm kin} + {\cal L}_{\rm M}^{\rm kin} + {\cal L}_{\rm BM} - V_{\rm NL}
\label{lagrangian}
\end{IEEEeqnarray}
where, ${\cal L}_B$ is the kinetic term for the baryons given as
\begin{IEEEeqnarray}{rCl}
{\cal L}_{\rm B}^{\rm kin} = \sum_{b \in {\rm B}}\bar\psi_{\rm b}(i\gamma^\mu\partial_\mu - m_{\rm b})\psi_{\rm b}. \label{lb}
\end{IEEEeqnarray}
Here, $\psi_{\rm b}$ and $m_{\rm b}$ correspond to the baryonic field and its bare mass, respectively. Similarl,y the kinetic term for the mesons is given by,
\begin{IEEEeqnarray}{rCl}
{\cal L}_{\rm M}^{\rm kin} &=& \frac{1}{2} \left[ \partial_{\mu}\sigma\partial^{\mu}\sigma - m_{\sigma}^2\sigma^2 \right] - \frac{1}{4}\Omega_{\mu\nu}\Omega^{\mu\nu} + \frac{1}{2} m_{\omega}^2\omega^2 \nonumber \\
&& - \frac{1}{4}{\rm \bf R}_{\mu\nu}{\rm \bf R}^{\mu\nu} + \frac{1}{2}m_{\rho}^2\rho_{\mu}\rho^{\mu} - \frac{1}{4}{\Phi}_{\mu\nu}{\Phi}^{\mu\nu}, \nonumber \\
&& \label{lmkin}
\end{IEEEeqnarray}
with $\Omega^{\mu\nu} = \partial^{\mu}\omega^{\nu} - \partial^{\nu}\omega^{\mu}$, ${\rm \bf R}^{\mu\nu} = \partial^{\mu}{\rm \bf \rho}^{\nu} - \partial^{\nu}{\rm \bf \rho}^{\mu}$ are the mesonic field strength tensors. The $\sigma$, $\omega$ and $\rho$ meson fields are denoted by $\sigma$, $\omega$ and $\rho$ and their masses are $m_{\sigma}$, $m_{\omega}$ and $m_{\rho}$, respectively. 
The Lagrangian ${\cal L}_{\rm BM}$ is the Lagrangian describing the baryon meson interactions having the form,
\begin{IEEEeqnarray}{rCl}
{\cal L}_{\rm BM} &=& \sum_{{\rm b}\ \in\ {\rm B}} \bar\psi_{\rm b}g_{\sigma {\rm b}}\sigma\psi_{\rm b} - \bar\psi_{\rm b}\gamma_{\mu}\left(g_{\omega {\rm b}}\omega^{\mu} + g_{\rho {\rm b}}{\bf \tau}_{\rm b}\cdot {\bf \rho}^{\mu} \right)\psi_{\rm b}, \nonumber \\
&&
\end{IEEEeqnarray}
where, ${\bf \tau}_{\rm b}$ is the Pauli matrix, and $g_{\alpha {\rm b}}$ for $\alpha \in \sigma, \omega^{\mu}, {\bf \rho}$ are the coupling constants of the baryons with the mesons. Similarly, $V_{\rm NL}$ describes the nonlinear interaction of mesons which is given by,
\begin{IEEEeqnarray}{rCl}
V_{\rm NL} &=& \frac{\kappa}{3!}(g_{\sigma {\rm N}}\sigma)^3 + \frac{\lambda}{4!}(g_{\sigma {\rm N}}\sigma)^4 - \frac{\xi_\omega}{4!}(g_{\omega {\rm N}}^2\omega_{\mu}\omega^{\mu})^2 \nonumber \\
&& - {\Lambda_{\omega\rho}}(g_{\omega {\rm N}}^2\omega_{\mu}\omega^{\mu})(g_{\rho {\rm N}}^2\rho_{\mu}\rho^{\mu}), 
\end{IEEEeqnarray}

In \ac{rmf} approximation, one replaces the meson fields by their expectation values which then act as classical fields in which baryons move $i.e.$ $\langle\sigma\rangle=\sigma_0$, $\langle \omega_\mu\rangle=\omega_0\delta_{\mu 0}$, $\langle \rho_\mu^a\rangle$ =$\delta_{\mu 0}\delta_{3}^a \rho_{3}^0$. The mesonic field equations of motion can be found by the Euler-Lagrange equations for the meson fields using the Lagrangian Eq. (\ref{lagrangian})
\begin{IEEEeqnarray}{rCl}
m_{\sigma}^2 \sigma_0 + \frac{\kappa}{2} g_{\sigma {\rm N}}^3 \sigma_0^2 + \frac{\lambda}{3!} g_{\sigma {\rm N}}^4 \sigma_0^3 &=& \sum_{\rm b} g_{\sigma {\rm b}}n_{\rm b}^s, \label{fieldeqns.sigma}
\end{IEEEeqnarray}
\begin{IEEEeqnarray}{rCl}
m_{\omega}^2 \omega_0 + \frac{\xi_\omega}{3!} g_{\omega {\rm N}}^4 \omega_0^3 + 2\Lambda_{\omega\rho} (g_{\rho {\rm N}} g_{\omega {\rm N}} \rho_0)^2 \omega_0 &=& \sum_{\rm b} g_{\omega {\rm b}}n_{\rm b}, \nonumber \\
& \label{fieldeqns.omega}
\end{IEEEeqnarray}
\begin{IEEEeqnarray}{rCl}
m_{\rho}^2 \rho_3^0 + 2\Lambda_{\omega\rho} (g_{\rho {\rm N}} g_{\omega {\rm N}} \omega_0)^2 \rho_0 &=& \sum_{\rm b} g_{\rho {\rm b}}I_{3{\rm b}}n_{\rm b}, \label{fieldeqns.rho}
\end{IEEEeqnarray}
where $I_{3{\rm b}}$ is the third component of the isospin of a given baryon. We have taken $I_{3 (n,p)} = \left(-\frac{1}{2}, \frac{1}{2}\right)$. The baryon density, $n_{\rm B}$, and scalar density, $n^s_{\rm b}$, at zero temperature are given as,
\begin{IEEEeqnarray}{rCl}
n_{\rm B} &=& \sum_{\rm b} \frac{\gamma k_{\rm Fb}^3}{6\pi^2} \equiv \sum_{\rm b} n_{\rm b}, \label{baryon.density}
\end{IEEEeqnarray}
and,
\begin{IEEEeqnarray}{rCl}
n^s_{\rm B} &=& \frac{\gamma}{(2 \pi)^3} \sum_{\rm b} \int_0^{k_{\rm Fb}}\frac{m^*_{\rm b}}{E(k)} d^3k \equiv \sum_{\rm b} n^s_{\rm b}, \label{baryon.scalar.density}
\end{IEEEeqnarray}
where $E(k) = \sqrt{m^{*}_{\rm b}{^2}+ k^2}$ is the single particle energy for nucleons with a medium-dependent mass given as 
\begin{equation}
m^{*}_{\rm b} = m_{\rm b} -g_{\sigma {\rm b}}\sigma_0. \label{effective.mass}
\end{equation}
Further, $k_{\rm Fb}=\sqrt{ \tilde{\mu}_{\rm b}{}^2 - m^{*}_{\rm b}{}^2}$ is the Fermi momenta of a baryon defined through an effective baryonic chemical potential, $\tilde{\mu}_{\rm b}$ given as,
\begin{equation}
\tilde{\mu}_{\rm b} = {\mu}_{\rm b} - g_{\omega {\rm b}}\omega_0 - g_{\rho {\rm b}}I_{\rm 3b}\rho_3^0. \label{effective.chemical.potential}
\end{equation}

Further, $\gamma=2$ corresponds to the spin degeneracy factor for nucleons and leptons and $\mu_l$ denotes the chemical potential for leptons. The energy density, $\epsilon_{\rm HP}$, within the \ac{rmf} model is given by
\begin{IEEEeqnarray}{rCl}
\epsilon_{\rm HP} &=& \frac{m^{*}_{\rm b}{}^4}{\pi^2}\sum_{\rm b} H(k_{\rm Fb}/m^{*}_{\rm b}) \nonumber
\\
&& + \frac{1}{2}m_{\sigma}^2\sigma_0^2 + \frac{1}{2} m_{\omega}^2\omega_0^2 + \frac{1}{2} m_{\rho}^2{\rho_{3}^0}^2 + \frac{\kappa}{3!}(g_{\sigma{\rm N}}\sigma_0)^3 \nonumber
\\
&& + \frac{\lambda}{4!}(g_{\sigma}\sigma_0)^4 + \frac{\xi_{\omega}}{8}(g_{\omega}\omega)^4 + 3{\Lambda_{\omega\rho}}(g_{\rho}g_{\omega}\rho_0 \omega_0)^2
\nonumber
\\
&& \label{energy.density.nm}
\end{IEEEeqnarray}

\noindent The pressure, $p_{\rm HP}$, can be found using the thermodynamic relation as
\begin{IEEEeqnarray}{rCl}
p_{\rm HP} &=& \sum_{\rm b} \mu_{\rm b} n_{\rm b} - \epsilon_{\rm HP}. \label{pressure.nm}
\end{IEEEeqnarray}

\noindent In Eq. (\ref{energy.density.nm}) we have introduced the function $H(z)$ which is given as
\begin{IEEEeqnarray}{rCl}
H(z) &=& \dfrac{1}{8} \left[z\sqrt{1+z^2}(1+2z^2)-\sinh^{-1}z \right], \label{function.h}
\end{IEEEeqnarray}
The \ac{ns} matter is charge neutral as well as $\beta$-equilibrated, i.e. $n \to p + e$ and $n \to p + \mu$, where $n$, $p$, $e$ and $\mu$ denote neutron, proton, electron and muon, respectively. We have taken the leptonic contribution to get the \ac{eos} for the core of \ac{ns}. The chemical potentials and number densities of constituents of \ac{nsm} are related by the following equations,
\begin{IEEEeqnarray}{rCl}
{\mu}_i = {\mu}_B &+& q_i {\mu}_E, \label{beta.equalibrium.rmf}
\\
\sum_{i=n,p,l} n_iq_i &=& 0,
\end{IEEEeqnarray}
where, $\mu_B$ and $\mu_E$ are the baryon and electric chemical potentials and $q_i$ is the charge of the $i^{th}$ particle.

\section{Neutron star properties \label{neutron_star_properties}}
The general static spherically-symmetric metric which describe the geometry of a static NS can be written as
\begin{IEEEeqnarray}{rCl}
ds^2 &=& e^{2\nu(r)} dt^2-e^{2\lambda(r)} dr^2-r^2 (d\theta^2+\sin^2\theta d\phi^2), \nonumber \label{metric} \\
\end{IEEEeqnarray}
where, $\nu(r)$ and $\lambda(r)$ are the metric functions. It is convenient to define the mass function, $m(r)$ in the favor of $\lambda$(r) as 
\begin{equation}
e^{2\lambda(r)} = \left(1-\frac{2m(r)}{r}\right)^{-1}.
\end{equation}
Starting from the line element, Eq. (\ref{metric}), one can obtain the equations governing the structure of spherical compact objects, the Tolmann-Oppenheimer-Volkoff (TOV) equations, as 
\begin{IEEEeqnarray}{rCl}
\frac{dp(r)}{dr} &=& -\left(\epsilon +p \right)\frac{d\nu }{dr}, \label{tov.pressure}
\\
\frac{dm(r)}{dr} &=& 4\pi r^2 \epsilon, \label{tov.mass}
\end{IEEEeqnarray}
\begin{IEEEeqnarray}{rCl}
\frac{d\nu(r)}{dr} &=& \frac{m+4 \pi r^3p}{r(r-2m)}. \label{tov.phi}
\end{IEEEeqnarray}
In the above set of equations $\epsilon(r)$, $p(r)$ are the energy density and the pressure, respectively. $m(r)$ is the mass of the compact star enclosed within a radius $r$. The boundary conditions $m(r=0) = 0$ and $p(r=0)=p_c$ and $p(r=R) = 0$, where $p_c$ is the central pressure lead to equilibrium configuration in combination with the \ac{eos} of NS matter, thus obtaining radius $R$ and mass $M = m(R)$ of NS for a given central pressure, $p_c$, or energy density, $\epsilon_c$. For a set of central energy densities $\epsilon_c$, one can obtain the mass-radius (M-R) curve. 

When a spherically symmetric \ac{ns} is subjected to the effect of a gravitational tidal field $\mathcal{E}_{ij}$, it develops a quadrupole moment $\mathcal{Q}_{ij}$ as a response to $\mathcal{E}_{ij}$. The parameter which relates $\mathcal{E}_{ij}$ and $\mathcal{Q}_{ij}$ quantities is the tidal deformability $\lambda = \frac{\mathcal{E}_{ij}}{\mathcal{Q}_{ij}}$ \cite{Hinderer:2009ca}. The parameter $\lambda$ can be related to a dimensionless number, Love number ($k_2$) as
\begin{IEEEeqnarray}{rCl}
\lambda &=& \frac{2}{3}k_2 R^5 \label{tidal}
\end{IEEEeqnarray}
and could be estimated by solving following differential equation along with \ac{tov} equations, \cite{Flores:2024hts}
\begin{IEEEeqnarray}{rCl}
\frac{dy}{dr} &=& -\frac{1}{r}\left(y^2 + y f_r + r^2 q_r\right), \label{tidal.ay}
\end{IEEEeqnarray}
where,
\begin{IEEEeqnarray}{rCl}
f_r &=& (1 - 4\pi r^2(\epsilon - p))\left(1 - \frac{2m}{r}\right)^{-1}, \nonumber \\
q_r &=& 4\pi\left(5\epsilon + 9p + (\epsilon + p) c_s^2 - \frac{6}{4\pi r^2}\right) \left(1 - \frac{2m}{r}\right)^{-1} \nonumber \\
      && - 4\left(\frac{d\nu(r)}{dr}\right)^2. \nonumber
\end{IEEEeqnarray}
The love number can be estimated by substituting the value of $y$ at the surface of a \ac{ns} into the following definition,
\begin{IEEEeqnarray}{rCl}
k_2 &=& \frac{8C^5}{5}\left(1 - 2C\right)^2 \left(2 + 2C(y - 1) - y\right)(T_0 + T_1 + T_2)^{-1} \label{love_number} \nonumber \\
&&
\end{IEEEeqnarray}
where,
\begin{IEEEeqnarray}{rCl}
T_0 &=& 2C\left(6 - 3y + 3C(5y - 8)\right), \nonumber \\
T_1 &=& 4C^3 \left(13 - 11y + C(3y - 2) + 2C^2(1 + y)\right), \nonumber \\
T_2 &=& 3\left(1 - 2C^2\right)\left(2 - y + 2C(y - 1)\right)\log\left(1 - 2C\right), \nonumber
\end{IEEEeqnarray}
and the compactness parameter is defined as $C=\frac{M}{R}$ while the dimensionless tidal deformability i.e. weighted tidal deformability is defined as 
\begin{IEEEeqnarray}{rCl}
\Lambda &=& \frac{2}{3}k_2 \left(\frac{R}{M}\right)^5. \label{tidal_weighted}
\end{IEEEeqnarray}

\section{Bayesian analysis} \label{sec:bayesian_analysis}
To fix the coupling constants, we use the Bayesian analysis \cite{Malik:2022zol}. It enables one to carry out a detailed statistical analysis of the parameters of a model for a given set of fit data \cite{Wesolowski:2015fqa, Furnstahl:2015rha,  Landry:2020vaw}. To a good approximation the \ac{eos} of nuclear matter can be decomposed into two parts (i) the \ac{eos} for symmetric nuclear matter $\epsilon(\rho,0)$ and (ii) a term involving the symmetry energy coefficient $S(\rho)$ and isospin asymmetry parameter $\delta$ ( $\delta=(\rho_n-\rho_p)/\rho$),
\begin{IEEEeqnarray}{rCl}
\epsilon(\rho,\delta) &\simeq & \epsilon(\rho,0)+S(\rho)\delta^2, \label{eq:eden}
\end{IEEEeqnarray}
where $\epsilon$ is the energy per nucleon at a given density $\rho$. We can recast the \ac{eos} in terms of various bulk nuclear matter properties of order $n$ at saturation density, $\rho_0$: (i) for the symmetric nuclear matter, the energy per nucleon $\epsilon_0=\epsilon(\rho_0,0)$ ($n=0$), the incompressibility coefficient $K_0$ ($n=2$), the skewness  $Q_0$ ($n=3$),  and  the kurtosis $Z_0$ ($n=4$), respectively, given by \cite{Malik:2022zol}
\begin{equation}
X_0^{(n)}=3^n \rho_0^n \left (\frac{\partial^n \epsilon(\rho, 0)}{\partial \rho^n}\right)_{\rho_0}, \qquad n=2,3,4; \label{x0}
\end{equation}
(ii) for the symmetry energy, the symmetry energy at saturation density $J_{\rm sym,0}$ ($n=0$), 
\begin{equation}
J_{\rm sym,0}= S(\rho_0), \quad S(\rho)=\frac{1}{2} \left (\frac{\partial^2 \epsilon(\rho,\delta)}{\partial\delta^2}\right)_{\delta=0},
\end{equation}
the slope $L_{\rm sym,0}$ ($n=1$), the curvature $K_{\rm sym,0}$ ($n=2$), the skewness $Q_{\rm sym,0}$ ($n=3$) and the kurtosis $Z_{\rm sym,0}$ ($n=4$) are defined as
\begin{equation}
X_{\rm sym,0}^{(n)} = 3^n \rho_0^n \left (\frac{\partial^n S(\rho)}{\partial \rho^n}\right )_{\rho_0} \qquad n=1,2,3,4. \label{xsym}
\end{equation}

In the Bayesian analysis, the basic rules of probabilistic inference are used to update the probability for a hypothesis under the available evidence according to Bayes' theorem. The posterior distributions of the model parameters $\theta$ in Bayes theorem can be written as \cite{Malik:2022zol}
\begin{equation}
P(\bm{\theta} |D ) =\frac{{\mathcal L } (D|\bm{\theta}) P(\bm {\theta })}{\mathcal Z},\label{eq:bt}
\end{equation}
where $\bm{\theta}$ and $D$ denote the set of model parameters and the fit data. $P(\bm {\theta })$ in Eq. (\ref{eq:bt}) is the prior for the  model parameters and $\mathcal Z$ is the evidence. The type of prior can be chosen with the preliminary knowledge of the model parameters. The $P(\bm{\theta} |D )$ is the joint posterior distribution of the parameters, $\mathcal L (D|\bm{\theta})$ is the likelihood function. We collect the experimental data values of the nuclear saturation properties in TABLE \ref{table-nmsp}.

\begin{table}[h!]
\centering
\caption{Nuclear saturation properties at saturation density $\rho_0 = 0.16\pm 0.02\ {\rm fm}^{-3}$, \ac{ns} mass and radius and corresponding numerical values \cite{sym16020215, Jiang:2022tzf}. These are the observations are used here in the Bayesian analysis to estimate the isoscalar-vector isovector-vector coupling strength. We have considered subluminal ($c_s^2 \le 1$) \ac{eos}s only. \label{table-nmsp}}
\begin{tabular}{|l|r|l|}
\hline
\multicolumn{1}{|c}{{\bf Nuclear saturation}} & \multicolumn{1}{|c}{{\bf Numerical}} & \multicolumn{1}{|c|}{{\bf References}}\\
\multicolumn{1}{|c}{{\bf properties}} & \multicolumn{1}{|c|}{{\bf values}} & \\
\hline \hline
{$E/A\ ({\rm MeV})$}        & $-16.5\pm 0.05$  & \cite{sym16020215} \\
{$K_0\ ({\rm MeV})$}        & $240\pm 20$      & \cite{Garg:2018uam,Shlomo:2006ole} \\
{$J_0\ ({\rm MeV})$}        & $32.5\pm 1.5$    & \cite{Oertel:2016bki} \\
{$L_0\ ({\rm MeV})$}        & $40\pm 20$       & \cite{Oertel:2016bki} \\
{$K_{\rm sym,0}\ ({\rm MeV})$}  & $-100\pm 50$ & \cite{Li:2020ass}     \\
{$M\ ({\rm M}_{\odot})$}    & $2.05\pm 0.04$   & \cite{Miller:2021qha, Dittmann:2024mbo} \\
{$R\ ({\rm km})$}           & $12.35\pm 0.75$  & \cite{Miller:2019cac, Riley:2019yda, Vinciguerra:2023qxq} \\
{$R\ ({\rm km})$}           & $12.45\pm 0.65$  & \cite{Miller:2021qha, Dittmann:2024mbo} \\
{$R\ ({\rm km})$}           & $11.45\pm 0.60$  & \cite{LIGOScientific:2017vwq, LIGOScientific:2018cki, De:2018uhw} \\
{$\Lambda\ ({\rm km})$}     & $190 \pm 85$     & \cite{LIGOScientific:2017vwq, LIGOScientific:2018cki, De:2018uhw} \\
\hline
\end{tabular}%
\end{table}%


\section{Results and discussions} \label{sec:results_and_discussions}
The original FSUGold parameter set \cite{Todd-Rutel:2005yzo} of \ac{rmf} model satisfies nuclear saturation properties and finite nuclei but it does not give the the maximum mass ($\ge M_{\odot}$) \ac{ns} constraint found in observations, PSR J0740+6620 and \cite{Miller:2021qha, Dittmann:2024mbo} PSR J0348+0432 \cite{Antoniadis:2013pzd}. Another group updates the FSUGold parameter set by lowering the $g_{\sigma}$, and $g_{\omega}$ couplings, $\xi$ and raising the $g_{\rho}$, and $\Lambda_{\omega\rho}$, yielding a stiffer high‑density \ac{eos} which keeps radii small and higher \ac{ns} mass and renaming their parameters as IU-FSU  \cite{Fattoyev:2010mx}. In the present study, we analyse the same using the Bayesian analysis and find a set of parameters which give \ac{eos} satisfying nuclear saturation properties and recent \ac{ns} observations. We choose some of the nuclear matter saturation properties such as binding energy per nucleon ($E/A$), incompressibility ($K_0$), symmetry energy ($J_0$), slope parameter ($L_0$) and incompressibility of symmetry energy ($K_{\rm sym, 0}$) at saturation density ($\rho_0$) to constrain \ac{eos} at low densities. Furthermore, to study the strong interactions of nuclear matter at high densities, we also consider astrophysical observations of \ac{ns}s such as mass ($M$), radius ($R$) and tidal deformability ($\Lambda$) obtained from the \ac{nicer} and \ac{gw} experiments. The numerical values of these experimental and observational data are collected in TABLE \ref{table-nmsp}. 

The Lagrangian in Eq. (\ref{lagrangian}) has various parameters, e.g. baryon masses, meson masses and various coupling parameters. We are considering here, $npe$ isospin symmetric matter with masses of neutron and proton as $m_n = m_p = 939\ {\rm MeV}$ and mass of electron is considered as $m_e = 0.511\ {\rm MeV}$. The masses of mesons are taken here as $m_{\sigma} = 491.5\ {\rm MeV}$, $m_{\omega} = 782.5\ {\rm MeV}$ and $m_{\rho} = 763.0\ {\rm MeV}$ respectively. We present the results of our Bayesian analysis focusing on constraining the \ac{eos}, which produces a larger \ac{ns} mass and satisfies nuclear saturation properties. The other coupling constants e.g. the Yukawa coupling constants: ($g_\sigma$, $g_\omega$, $g_\rho$), the self-couplings: ($\kappa$, $\lambda$, $\zeta$) and the most important term of the present work, the crossed coupling constant: ($\Lambda_{\omega\rho}$) are taken to vary freely and constrained by the nuclear saturation properties and astrophysical observations using the Bayesian analysis. We sampled about {\em 60 millions} of parameters and found about {\em 20000} parameters of the model which satisfy the nuclear matter properties, the \ac{pnm} \ac{eos} calculated from a precise N$^3$LO calculation in \ac{ceft} and the lowest bound of \ac{ns} observational maximum mass. In the following section, we discuss the obtained results.

\begin{figure}[ht]
    \centering
    \includegraphics[width=0.95\linewidth]{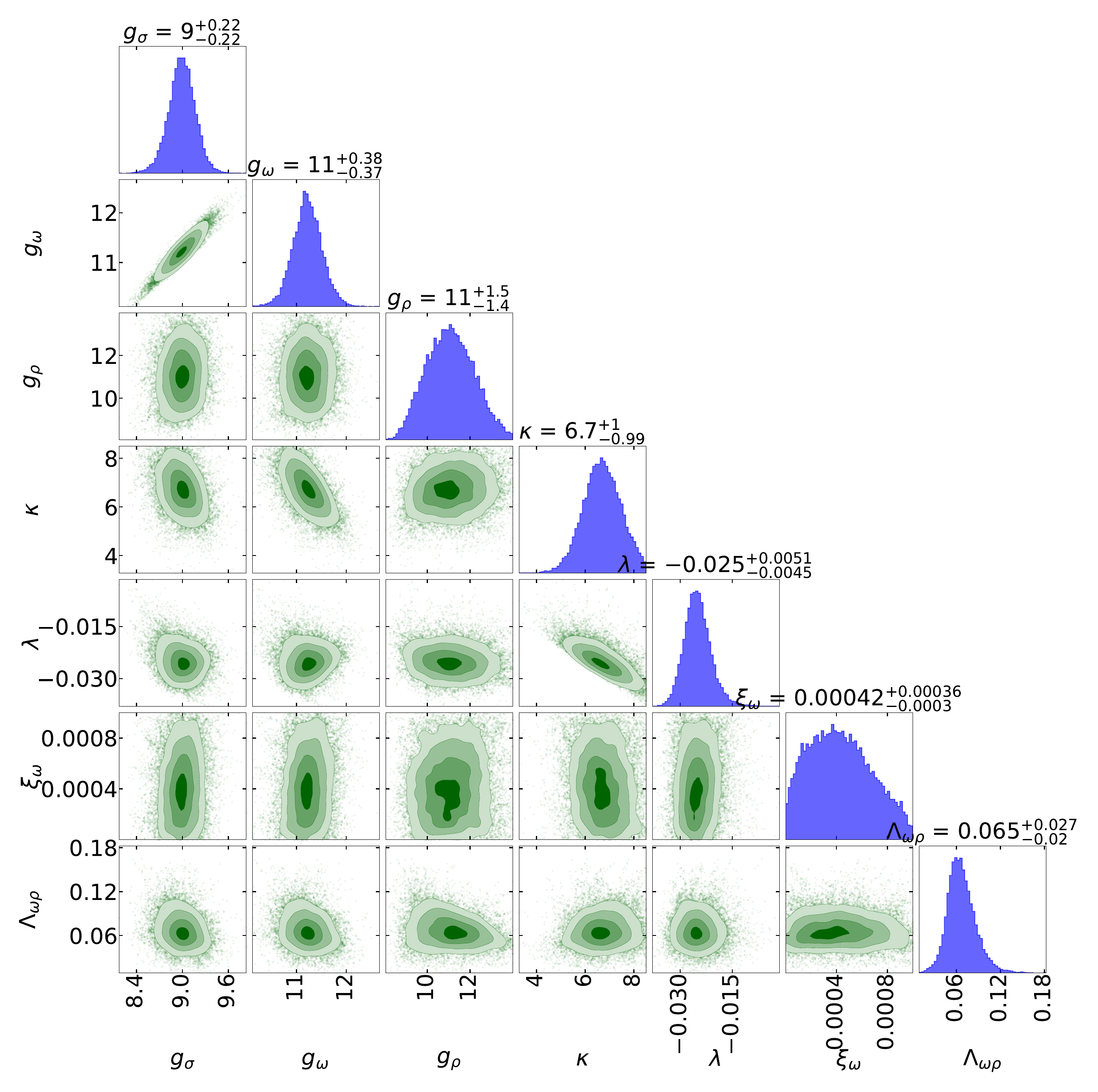}
    \caption{The posterior distribution of the coupling parameters ($g_{\sigma}$, $g_{\omega}$, $g_{\rho}$, $\kappa$, $\lambda$, $\xi_{\omega}$, $\Lambda_{\omega\rho}$). The peak in each histogram shows the best-fit central value of a corresponding parameter.}
    \label{fig:cplngs}
\end{figure}
\subsection{Nuclear matter saturation properties} \label{sec:nuclear_matter_saturation_properties}
In this section, we analyse the \ac{eos} for the present case as described in Eqs. (\ref{energy.density.nm}) and (\ref{pressure.nm}). Before going to discuss the \ac{eos} sets, we first discuss the coupling parameters found from the Bayesian analysis. We consider, here, the uniform distributions for each coupling parameter and set a moderate range for $g_{\sigma}$. We allow the cross-coupling constant,  $\Lambda_{\omega\rho}$, in a range (0,0.2). We find that the larger value of $\Lambda_{\omega\rho}$ coupling gives a larger mass without changing the nuclear saturation properties. Here, we get $\Lambda_{\omega\rho} = 0.065^{+0.027}_{-0.02}$. In FIG. \ref{fig:cplngs}, we display the posterior distributions for all  coupling constants ($g_{\sigma}$, $g_{\omega}$, $g_{\rho}$, $\kappa$, $\lambda$, $\xi_{\omega}$, $\Lambda_{\omega\rho}$) obtained after the Bayesian analysis. The diagonal panels show the one-dimensional marginalized distributions for each coupling constant, with the central values indicated by the peaks in each histograms.

\begin{figure}[ht]
    \centering
    \includegraphics[width=0.95\linewidth]{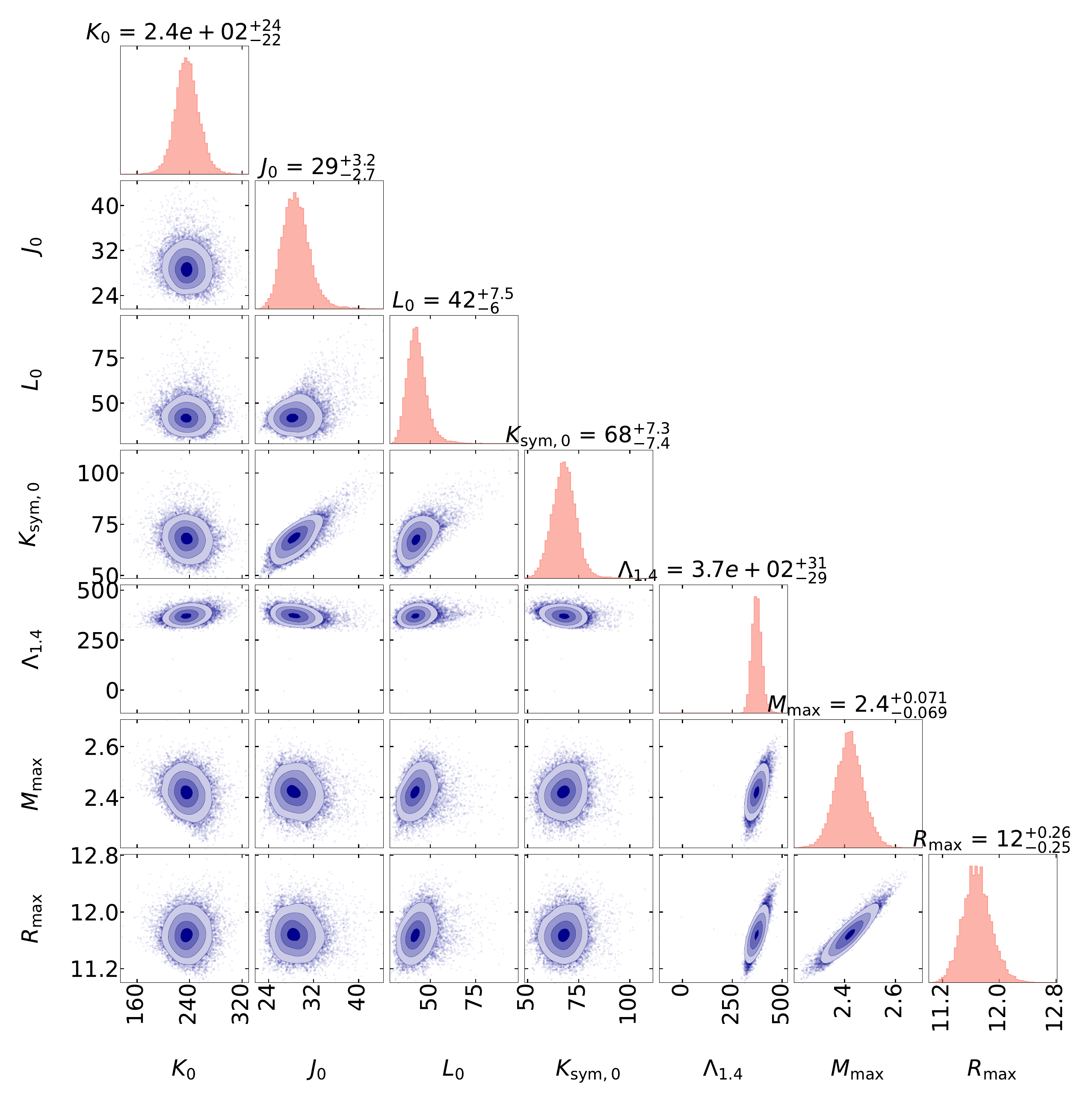}
    \caption{Distribution of the nuclear saturation properties, the maximum masses and corresponding radii of \ac{ns}s which are evaluated for all the \ac{eos} sets for which the coupling parameters are obtained as above. The peak in each histogram shows the best-fit central value of a corresponding parameter.}
    \label{fig:snmps}
\end{figure}
Next, we discuss the best-fit values of these quantities and \ac{ns} properties obtained in the present study for all the \ac{eos} sets. We see from the FIG. \ref{fig:cplngs} that the couplings $\lambda$ and $\Lambda_{\omega\rho}$ are more sensitive to produce high mass \ac{ns}s. The $\lambda$ becomes negative and the corresponding value of $\Lambda_{\omega\rho}$ is large as compared to the FSUGold \cite{Todd-Rutel:2005yzo} and IU-FSU \cite{Fattoyev:2010mx} parameter sets. With these scenarios, we can accommodate very large \ac{ns} without considering dark matter, quark matter or another form of exotic matter as taken in the Ref. \cite{Dey:2024vsw}. Our main focus of the present work is to determine the isovector-vector and isoscalar-vector coupling ($\Lambda_{\omega \rho}$). Here, we see that its value found to be $\Lambda_{\omega \rho} \in (0.009, 0.182)$ with the mean $0.067$ and standard deviation $\sigma = 0.019$. As we see that as the value of $\Lambda_{\omega\rho}$ increases, the \ac{eos} become stiffer and produce heavier \ac{ns}. In TABLE \ref{table-cplg_sel20}, we give a few sets of the coupling parameters and corresponding \ac{ns} properties e.g. maximum mass and its radius. From Sr. No. 1 to 25 (26-35), we give sets of coupling parameters which produce high mass $\geq 2{\rm M}_{\odot}$ ($\geq 2.5{\rm M}_{\odot}$) \ac{ns}s. The full sets of coupling parameters and corresponding equations of states, and \ac{ns}s macroscopic parameters could be found in  \href{https://u.pcloud.link/publink/show?code=kZPL3m5Zc5a0w45VVl8Jhk2ragtDFh3l2JM7}{EOS DATA SET}.


\begin{table*}[htbp]
  \centering
  \caption{The coupling parameters as taken in Eq. (\ref{lagrangian}) and corresponding the maximum mass of \ac{ns}s and its radii obtained for the respected parameter set. From Sr No. 1 to 25 are a few parameters sets which give \ac{ns} of mass less than $2.5M_{\odot}$ while after 25 to 35 the same which give \ac{ns} of mass more than $2.5M_{\odot}$.}
    \begin{tabular}{|c|c|c|c|c|c|c|c|c|c|c|c|}
    \hline
    \multicolumn{1}{|c}{Sr. No.} & \multicolumn{1}{|c}{$g_{\sigma}$} & \multicolumn{1}{|c}{$g_{\omega}$} & \multicolumn{1}{|c}{$g_{\rho}$} & \multicolumn{1}{|c}{$\kappa$} & \multicolumn{1}{|c}{$\lambda$} & \multicolumn{1}{|c}{$\xi_{\omega}$} & \multicolumn{1}{|c}{$\Lambda_{\omega\rho}$} & \multicolumn{1}{|c}{$M_{\rm Max}$} & \multicolumn{1}{|c}{$R_{\rm Max}$} & \multicolumn{1}{|c}{$R_{\rm 1.4}$} & \multicolumn{1}{|c|}{$\Lambda_{\rm 1.4}$} \\
    \multicolumn{1}{|c}{} & \multicolumn{1}{|c}{} & \multicolumn{1}{|c}{} & \multicolumn{1}{|c}{} & \multicolumn{1}{|c}{(${\rm MeV}^{-1}$)} & \multicolumn{1}{|c}{} & \multicolumn{1}{|c}{} & \multicolumn{1}{|c}{} & \multicolumn{1}{|c}{($M_{\odot}$)} & \multicolumn{1}{|c}{(km)} & \multicolumn{1}{|c}{(km)} & \multicolumn{1}{|c|}{} \\
    \hline\hline
    1     & 8.8713 & 10.9532 & 9.3675 & 7.0597 & -0.0207 & 0.00083 & 0.0975 & 2.367 & 11.65 & 12.76 & 402.29 \\
    2     & 8.8603 & 10.9990 & 9.7837 & 6.7170 & -0.0215 & 0.00079 & 0.1027 & 2.376 & 11.61 & 12.62 & 388.02 \\
    3     & 8.9020 & 11.1542 & 11.5907 & 6.0359 & -0.0190 & 0.00013 & 0.0663 & 2.426 & 11.78 & 12.88 & 394.96 \\
    4     & 8.9239 & 11.1001 & 11.8025 & 6.3768 & -0.0199 & 0.00013 & 0.0584 & 2.414 & 11.73 & 12.87 & 387.72 \\
    5     & 8.9648 & 11.0263 & 12.4120 & 7.1295 & -0.0213 & 0.00064 & 0.0859 & 2.385 & 11.68 & 12.71 & 394.12 \\
    6     & 8.7868 & 10.8539 & 9.5626 & 6.9781 & -0.0226 & 0.00078 & 0.0954 & 2.349 & 11.49 & 12.54 & 368.61 \\
    7     & 8.8738 & 10.9554 & 9.8918 & 6.8644 & -0.0201 & 0.00069 & 0.0936 & 2.374 & 11.64 & 12.70 & 396.80 \\
    8     & 9.3115 & 11.8281 & 11.9231 & 5.0595 & -0.0180 & 0.00022 & 0.0684 & 2.556 & 12.25 & 13.09 & 450.23 \\
    9     & 9.2249 & 11.6304 & 11.9783 & 5.8591 & -0.0221 & 0.00049 & 0.0628 & 2.496 & 12.01 & 12.94 & 410.92 \\
    10    & 8.9425 & 11.1825 & 10.9965 & 5.9096 & -0.0178 & 0.00038 & 0.0902 & 2.430 & 11.82 & 12.80 & 408.90 \\
    11    & 8.6797 & 10.6157 & 11.6208 & 8.2200 & -0.0278 & 0.00057 & 0.0922 & 2.303 & 11.28 & 12.43 & 335.71 \\
    12    & 8.7026 & 10.7013 & 10.9124 & 7.4852 & -0.0253 & 0.00041 & 0.0923 & 2.330 & 11.34 & 12.41 & 343.14 \\
    13    & 8.5965 & 10.5317 & 10.7229 & 7.6280 & -0.0205 & 0.00059 & 0.0692 & 2.290 & 11.35 & 12.63 & 353.50 \\
    14    & 9.1644 & 11.5661 & 10.3930 & 5.7011 & -0.0214 & 0.00058 & 0.1136 & 2.492 & 11.99 & 12.77 & 421.03 \\
    15    & 8.5241 & 10.4743 & 12.2543 & 7.3646 & -0.0215 & 0.00052 & 0.0507 & 2.271 & 11.25 & 12.62 & 319.51 \\
    16    & 8.5905 & 10.5257 & 10.6586 & 7.5855 & -0.0199 & 0.00059 & 0.0676 & 2.289 & 11.35 & 12.65 & 354.76 \\
    17    & 8.9128 & 11.2910 & 11.6983 & 5.3915 & -0.0190 & 0.00021 & 0.0808 & 2.453 & 11.83 & 12.82 & 391.42 \\
    18    & 8.7897 & 10.9264 & 10.4779 & 6.8740 & -0.0241 & 0.00048 & 0.0881 & 2.369 & 11.49 & 12.53 & 360.64 \\
    19    & 8.5827 & 10.4978 & 10.6712 & 7.6872 & -0.0201 & 0.00059 & 0.0652 & 2.283 & 11.34 & 12.64 & 352.22 \\
    20    & 8.5007 & 10.4433 & 12.2427 & 7.4161 & -0.0217 & 0.00052 & 0.0498 & 2.264 & 11.23 & 12.61 & 315.84 \\
    21    & 9.1207 & 11.4099 & 10.7108 & 5.8682 & -0.0179 & 0.00057 & 0.0975 & 2.466 & 12.00 & 12.98 & 438.73 \\
    22    & 9.1381 & 11.4715 & 12.9433 & 5.9511 & -0.0197 & 0.00006 & 0.0317 & 2.470 & 12.03 & 13.41 & 410.84 \\
    23    & 8.6868 & 10.8152 & 10.0539 & 6.6290 & -0.0219 & 0.00040 & 0.1013 & 2.356 & 11.46 & 12.49 & 360.62 \\
    24    & 8.9718 & 11.1986 & 11.2855 & 6.4649 & -0.0235 & 0.00036 & 0.0661 & 2.421 & 11.70 & 12.70 & 375.18 \\
    25    & 9.0128 & 11.2762 & 11.1130 & 6.0471 & -0.0205 & 0.00045 & 0.0691 & 2.437 & 11.82 & 12.80 & 396.74 \\
    \hline\hline
    26    & 9.3115 & 11.8281 & 11.9231 & 5.0595 & -0.0180 & 0.00022 & 0.0684 & 2.556 & 12.25 & 13.09 & 450.23 \\
    27    & 9.3156 & 11.5959 & 11.5200 & 7.1096 & -0.0294 & 0.00028 & 0.0937 & 2.501 & 11.90 & 12.70 & 393.26 \\
    28    & 9.3267 & 11.8577 & 11.4267 & 5.0352 & -0.0177 & 0.00028 & 0.0698 & 2.558 & 12.28 & 13.14 & 459.39 \\
    29    & 9.3255 & 11.6096 & 11.5475 & 7.0961 & -0.0293 & 0.00027 & 0.0926 & 2.503 & 11.91 & 12.70 & 395.07 \\
    30    & 9.1797 & 11.6095 & 10.7258 & 5.6910 & -0.0217 & 0.00029 & 0.0766 & 2.507 & 12.02 & 12.88 & 417.95 \\
    31    & 9.4689 & 11.7905 & 11.7747 & 7.4416 & -0.0335 & 0.00016 & 0.0914 & 2.543 & 11.93 & 12.56 & 371.19 \\
    32    & 9.1360 & 11.5808 & 10.5147 & 5.5780 & -0.0210 & 0.00023 & 0.0789 & 2.505 & 12.02 & 12.89 & 419.64 \\
    33    & 9.3180 & 11.5934 & 11.5192 & 7.0985 & -0.0294 & 0.00028 & 0.0939 & 2.501 & 11.89 & 12.66 & 391.90 \\
    34    & 9.4432 & 11.7961 & 10.2707 & 6.9333 & -0.0301 & 0.00066 & 0.0623 & 2.514 & 11.98 & 12.74 & 384.96 \\
    35    & 9.4255 & 11.7774 & 10.2801 & 6.9037 & -0.0299 & 0.00065 & 0.0627 & 2.512 & 11.95 & 12.74 & 384.56 \\
    \hline
    \end{tabular}%
  \label{table-cplg_sel20}%
\end{table*}%


In FIG. \ref{fig:snmps}, we display the distribution of the nuclear saturation properties along with the maximum mass and respective radii of \ac{ns} evaluated for the \ac{eos} corresponding to a coupling parameter set as given in FIG. \ref{fig:cplngs}. We also display the central values of each nuclear saturation property and mass-radius duo along with their 90\% \ac{ci} shown as the vertical lines in the figure. We find that the values of nuclear saturation density $\rho = 0.151 \pm 0.004\ {\rm fm}^{-3}$, the binding energy per nucleon $BE = -15.943 \pm 0.280\ {\rm MeV}$, the incompressibility $K_0 = 240^{+24}_{-22}\ {\rm MeV}$, the symmetry energy $J_0 = 29^{+3.2}_{-2.7}\ {\rm MeV}$, the slope parameter $L_0 = 42^{+7.5}_{-6}\ {\rm MeV}$, the incompressibility of symmetry energy $K_{\rm sym, 0} = 68^{+7.3}_{-7.4}\ {\rm MeV}$, the maximum mass of \ac{ns}s $M_{\rm Max} = 2.4^{+0.07}_{-0.07}\ {\rm M}_{\odot}$, corresponding radius $R_{\rm Max} = 12^{+0.26}_{-0.25}\ {\rm km}$ and the tidal deformability of $M = 1.4\ {\rm M}_{\odot}$ $\Lambda_{\rm 1.4} = 370^{+31}_{-29}\ {\rm km}$.

\begin{table*}[htbp]
  \centering
  \caption{The domains of the nuclear saturation properties along with the other statistical quantities like Mean, standard deviation ($\sigma$), minima and maxima. \label{table-nprp_seldf}}
    \begin{tabular}{|c|c|c|c|c|c|c|c|c|c|}
    \hline
    Parameter & \multicolumn{1}{c|}{Mean} & \multicolumn{1}{c|}{Std} & \multicolumn{1}{c|}{$1\sigma$} & \multicolumn{1}{c|}{$2\sigma$} & \multicolumn{1}{c|}{$3\sigma$} \\
    \hline\hline
    $\rho_0\ ({\rm fm}^{-3})$ & 0.158 & 0.005 & (0.146, 0.164) & (0.148, 0.164) & (0.154, 0.164)  \\
    $BE\ ({\rm MeV})$  & -15.989 & 0.268 & (-16.497, -15.502) & (-16.466, -15.533) & (-16.298, -15.683)  \\
    $K_0\ ({\rm MeV})$ & 236.692 & 18.789 & (170.224, 304.694) & (199.321, 275.810) & (219.349, 254.486)  \\
    $J_0\ ({\rm MeV})$ & 28.985 & 2.476 & (23.106, 40.829) & (24.700, 34.557) & (26.625, 31.229) \\
    $L_0\ ({\rm MeV})$ & 42.854 & 6.014 & (29.941, 77.850) & (33.307, 57.324) & (37.484, 47.697) \\
    $K_{\rm sym,0}\ ({\rm MeV})$ & 67.974 & 6.011 & (51.483, 94.359) & (56.459, 80.424) & (62.161, 73.528) \\
    \hline\hline
    $M_{\rm Max}\ ({\rm M}_{\odot})$ & 2.420 & 0.055 & (2.245,  2.599) & (2.309, 2.532) & (2.366, 2.473) \\
    $R_{\rm Max}\ ({\rm km})$ & 11.690 & 0.204 & (11.140, 12.415) & (11.300, 12.125) & (11.490, 11.885)  \\
    $R_{\rm 1.4}\ ({\rm km})$ & 12.718 & 0.178 & (12.265, 13.557) & (12.396, 13.109) & (12.550, 12.877)  \\
    $\Lambda_{\rm 1.4}$ & 372.931 & 24.914 & (308.004, 464.866) & (326.988, 426.501) & (348.934, 396.408) \\
    \hline
    \end{tabular}%
  \label{tab:addlabel}%
\end{table*}%

In TABLE \ref{table-nprp_seldf}, we give the ranges of all the saturation quantities along with other statistical quantities for the distribution, e.g. mean, standard deviation, minima and maxima. Here, we see that the minima(maxima) of the saturation density is $\rho_0 = 0.146(0.158)\ {\rm fm}^{-3}$ while the central value of the distribution is $\rho_0 = 0.150\ {\rm fm}^{-3}$, which is in agreement with various recent studies (TABLE \ref{table-nmsp}). The binding energy per baryon is found to be $\in (-16.43, -15.54)\ {\rm MeV}$ with a standard deviation of $0.28\ {\rm MeV}$. This is also in very good agreement (TABLE \ref{table-nmsp}). The other nuclear saturation properties $K_0$, $J_0$, $L_0$, and $K_{\rm sym, 0}$ are also found with very small errors to be in good agreement with the experiments (TABLE \ref{table-nmsp}). We discuss the mass ($M$) and radius ($R$) in the following sections. In TABLE \ref{table-nprp_sel20}, we give the nuclear saturation properties, \ac{ns} mass and its radius corresponding to a few coupling parameter sets as found in TABLE \ref{table-cplg_sel20}.

\begin{table*}[htbp]
  \centering
  \caption{The nuclear saturation properties and the \ac{ns}s macroscopic properties corresponding to the coupling parameters collected in the TABLE \ref{table-cplg_sel20} \label{table-nprp_sel20}.}
    \begin{tabular}{|c|c|c|c|c|c|c|c|c|c|c|}
    \hline
    \multicolumn{1}{|c}{Sr. No.} & \multicolumn{1}{|c}{$\rho_o$} & \multicolumn{1}{|c}{$BE$} & \multicolumn{1}{|c}{$K_0$} & \multicolumn{1}{|c}{$J_0$} & \multicolumn{1}{|c}{$L_0$} & \multicolumn{1}{|c}{$K_{\rm sym,0}$} & \multicolumn{1}{|c}{$M_{\rm Max}$} & \multicolumn{1}{|c}{$R_{\rm Max}$} & \multicolumn{1}{|c}{$R_{\rm 1.4}$} & \multicolumn{1}{|c|}{$\Lambda_{\rm 1.4}$} \\
    \multicolumn{1}{|c}{} & \multicolumn{1}{|c}{(${\rm fm}^{\rm -3}$)} & \multicolumn{1}{|c}{(MeV)} & \multicolumn{1}{|c}{(MeV)} & \multicolumn{1}{|c}{(MeV)} & \multicolumn{1}{|c}{(MeV)} & \multicolumn{1}{|c}{(MeV)} & \multicolumn{1}{|c}{($M_{\odot}$)} & \multicolumn{1}{|c}{(km)} & \multicolumn{1}{|c}{(km)} & \multicolumn{1}{|c|}{} \\
    \hline\hline
    1     & 0.148 & -15.757 & 250.933 & 24.345 & 39.414 & 52.684 & 2.367 & 11.65 & 12.76 & 402.29 \\
    2     & 0.154 & -15.963 & 252.791 & 24.862 & 39.608 & 57.156 & 2.376 & 11.61 & 12.62 & 388.02 \\
    3     & 0.152 & -15.639 & 258.497 & 28.616 & 37.461 & 61.382 & 2.426 & 11.78 & 12.88 & 394.96 \\
    4     & 0.152 & -16.295 & 260.825 & 29.948 & 38.221 & 63.376 & 2.414 & 11.73 & 12.87 & 387.72 \\
    5     & 0.148 & -16.482 & 256.437 & 27.059 & 31.277 & 55.198 & 2.385 & 11.68 & 12.71 & 394.12 \\
    6     & 0.158 & -16.422 & 255.494 & 25.654 & 41.530 & 60.871 & 2.349 & 11.49 & 12.54 & 368.61 \\
    7     & 0.150 & -16.291 & 257.637 & 25.157 & 38.736 & 54.881 & 2.374 & 11.64 & 12.70 & 396.80 \\
    8     & 0.152 & -16.493 & 270.299 & 27.554 & 39.493 & 61.982 & 2.556 & 12.25 & 13.09 & 450.23 \\
    9     & 0.154 & -16.121 & 243.303 & 28.710 & 39.516 & 64.799 & 2.496 & 12.01 & 12.94 & 410.92 \\
    10    & 0.154 & -16.304 & 273.643 & 26.146 & 37.564 & 59.003 & 2.430 & 11.82 & 12.80 & 408.90 \\
    11    & 0.156 & -15.644 & 234.349 & 27.419 & 33.815 & 60.631 & 2.303 & 11.28 & 12.43 & 335.71 \\
    12    & 0.162 & -16.305 & 251.786 & 27.302 & 37.945 & 65.203 & 2.330 & 11.34 & 12.41 & 343.14 \\
    13    & 0.152 & -15.526 & 251.651 & 28.628 & 39.547 & 60.511 & 2.290 & 11.35 & 12.63 & 353.50 \\
    14    & 0.158 & -16.336 & 253.558 & 24.500 & 42.999 & 61.169 & 2.492 & 11.99 & 12.77 & 421.03 \\
    15    & 0.162 & -15.906 & 261.543 & 34.053 & 40.202 & 76.687 & 2.271 & 11.25 & 12.62 & 319.51 \\
    16    & 0.152 & -15.518 & 253.282 & 28.738 & 40.205 & 60.782 & 2.289 & 11.35 & 12.65 & 354.76 \\
    17    & 0.162 & -15.723 & 269.984 & 27.810 & 39.600 & 67.607 & 2.453 & 11.83 & 12.82 & 391.42 \\
    18    & 0.160 & -15.757 & 246.011 & 26.840 & 40.125 & 64.047 & 2.369 & 11.49 & 12.53 & 360.64 \\
    19    & 0.152 & -15.614 & 253.254 & 29.063 & 40.869 & 61.361 & 2.283 & 11.34 & 12.64 & 352.22 \\
    20    & 0.162 & -15.830 & 259.154 & 34.304 & 40.756 & 77.200 & 2.264 & 11.23 & 12.61 & 315.84 \\
    21    & 0.148 & -16.457 & 267.599 & 24.718 & 37.242 & 53.815 & 2.466 & 12.00 & 12.98 & 438.73 \\
    22    & 0.146 & -15.807 & 251.198 & 35.506 & 45.065 & 68.833 & 2.470 & 12.03 & 13.41 & 410.84 \\
    23    & 0.160 & -15.637 & 252.086 & 25.832 & 39.933 & 62.020 & 2.356 & 11.46 & 12.49 & 360.62 \\
    24    & 0.158 & -16.282 & 246.758 & 28.880 & 40.597 & 66.844 & 2.421 & 11.70 & 12.70 & 375.18 \\
    25    & 0.156 & -16.494 & 262.946 & 28.129 & 40.352 & 64.239 & 2.437 & 11.82 & 12.80 & 396.74 \\    \hline\hline
    26    & 0.152 & -16.493 & 270.299 & 27.554 & 39.493 & 61.982 & 2.556 & 12.25 & 13.09 & 450.23 \\
    27    & 0.154 & -16.117 & 203.976 & 25.633 & 39.438 & 60.723 & 2.501 & 11.90 & 12.70 & 393.26 \\
    28    & 0.150 & -16.266 & 267.873 & 26.910 & 40.188 & 59.905 & 2.558 & 12.28 & 13.14 & 459.39 \\
    29    & 0.152 & -16.109 & 201.854 & 25.536 & 38.690 & 59.290 & 2.503 & 11.91 & 12.70 & 395.07 \\
    30    & 0.156 & -15.925 & 246.856 & 26.733 & 42.704 & 63.568 & 2.507 & 12.02 & 12.88 & 417.95 \\
    31    & 0.158 & -15.940 & 155.449 & 26.054 & 42.997 & 66.095 & 2.543 & 11.93 & 12.56 & 371.19 \\
    32    & 0.156 & -15.592 & 249.133 & 26.442 & 42.886 & 63.001 & 2.505 & 12.02 & 12.89 & 419.64 \\
    33    & 0.154 & -16.290 & 203.505 & 25.634 & 39.492 & 60.759 & 2.501 & 11.89 & 12.66 & 391.90 \\
    34    & 0.160 & -16.410 & 183.986 & 28.017 & 49.139 & 71.026 & 2.514 & 11.98 & 12.74 & 384.96 \\
    35    & 0.160 & -16.414 & 186.147 & 27.997 & 48.918 & 70.856 & 2.512 & 11.95 & 12.74 & 384.56 \\
    \hline
    \end{tabular}%
  \label{tab:addlabel}%
\end{table*}%


\begin{figure}
    \centering
    \includegraphics[width=0.95\linewidth]{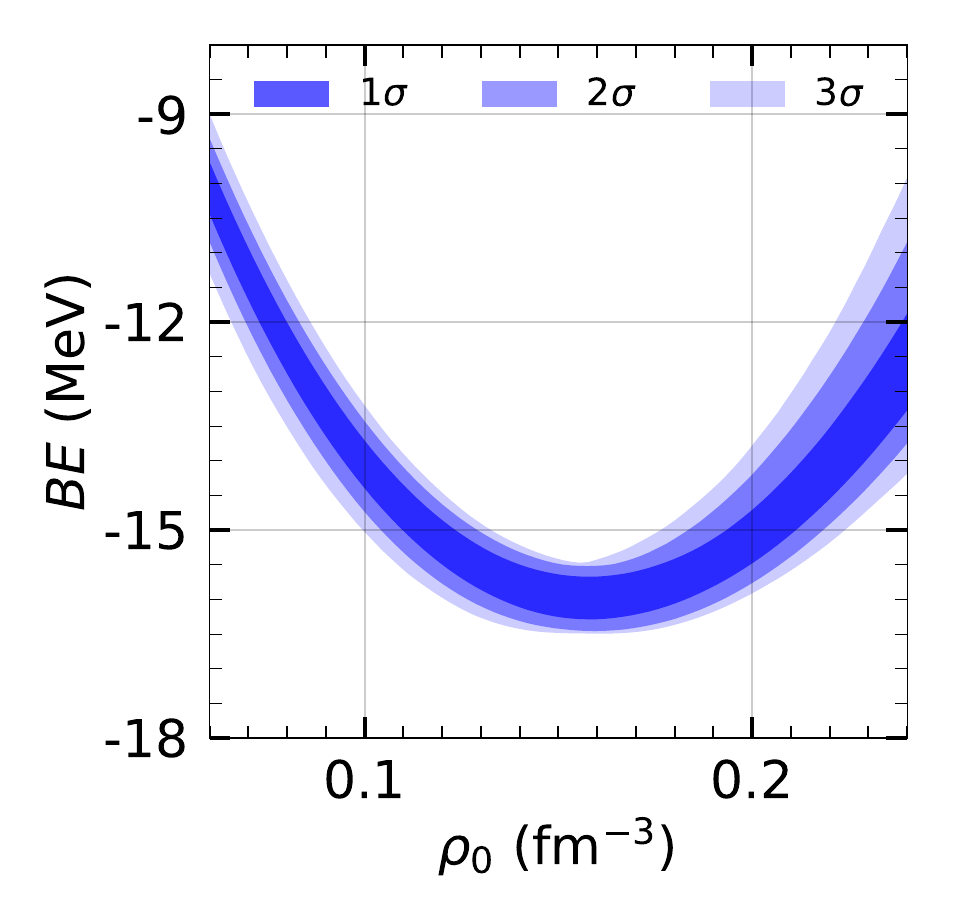}
    \includegraphics[width=0.95\linewidth]{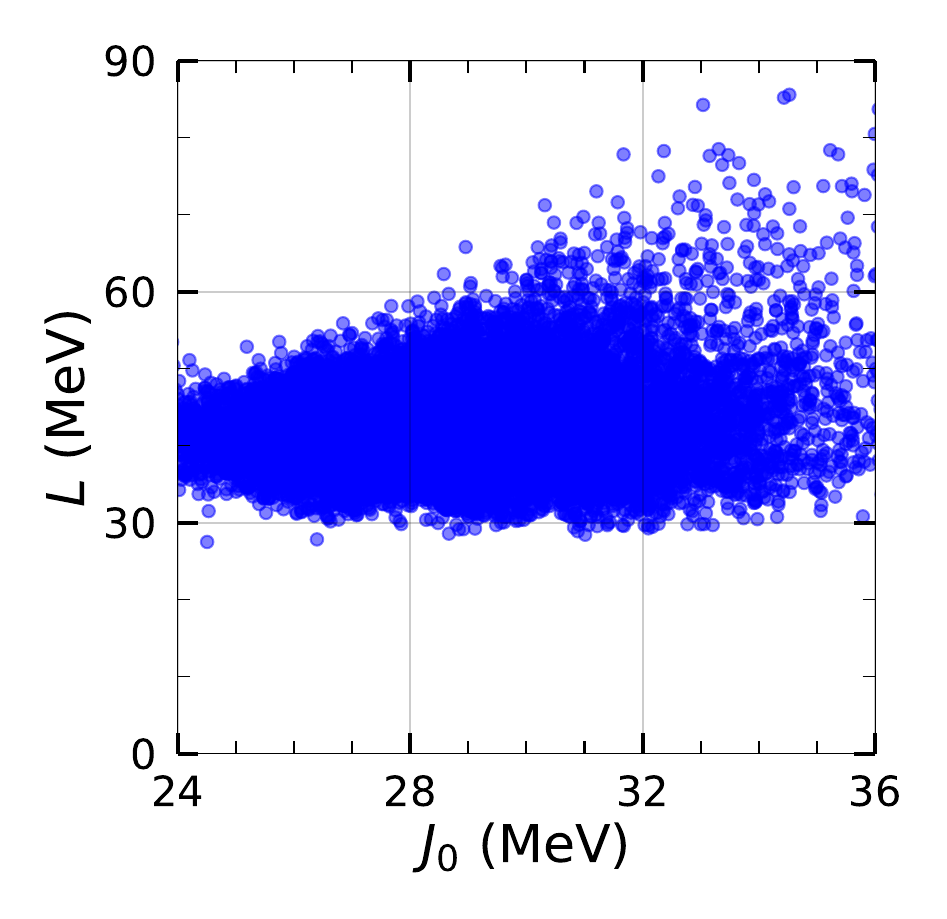}
    \caption{The variation of binding energy per baryon ($E/A$) as a function of nuclear density ($\rho_0$) in the upper panel. The different shades of blues show the $1\sigma$, $2\sigma$ and $3\sigma$ \ac{ci}s. In the lower panel, a scatter plot in between the slope parameter $L_0\ ({\rm MeV})$ and the symmetry energy $J_0\ ({\rm MeV})$. Each single point corresponds to a single \ac{eos}.}
    \label{fig:snmps_ben_and_snmps_ls0}
\end{figure}
In FIG. \ref{fig:snmps_ben_and_snmps_ls0}, we plot the 90\% \ac{ci} region for the binding energy per nucleon of symmetric matter as a function of baryon number density. In the top panel, we see that the minima of the binding energy per nucleon are around 0.145 to 0.16 ${\rm fm}^{-3}$ at which the pressure of the symmetric matter is zero, i.e. it defines the nuclear matter saturation density. At the nuclear saturation densities, the binding energy per baryon is found to be $BE \in (-16.43, -15.54)\ {\rm MeV}$, which is consistent with other theoretical and experimental constraints \cite{Dutra:2014qga}. In the bottom panel, we plot the correlation between the slope parameter ($L_0$) and the symmetry energy ($J_0$). Each dot in this scatter plot corresponds to a single \ac{eos} set obtained for each parameter set. 

\begin{figure*}
    \centering
    \includegraphics[width=0.45\linewidth]{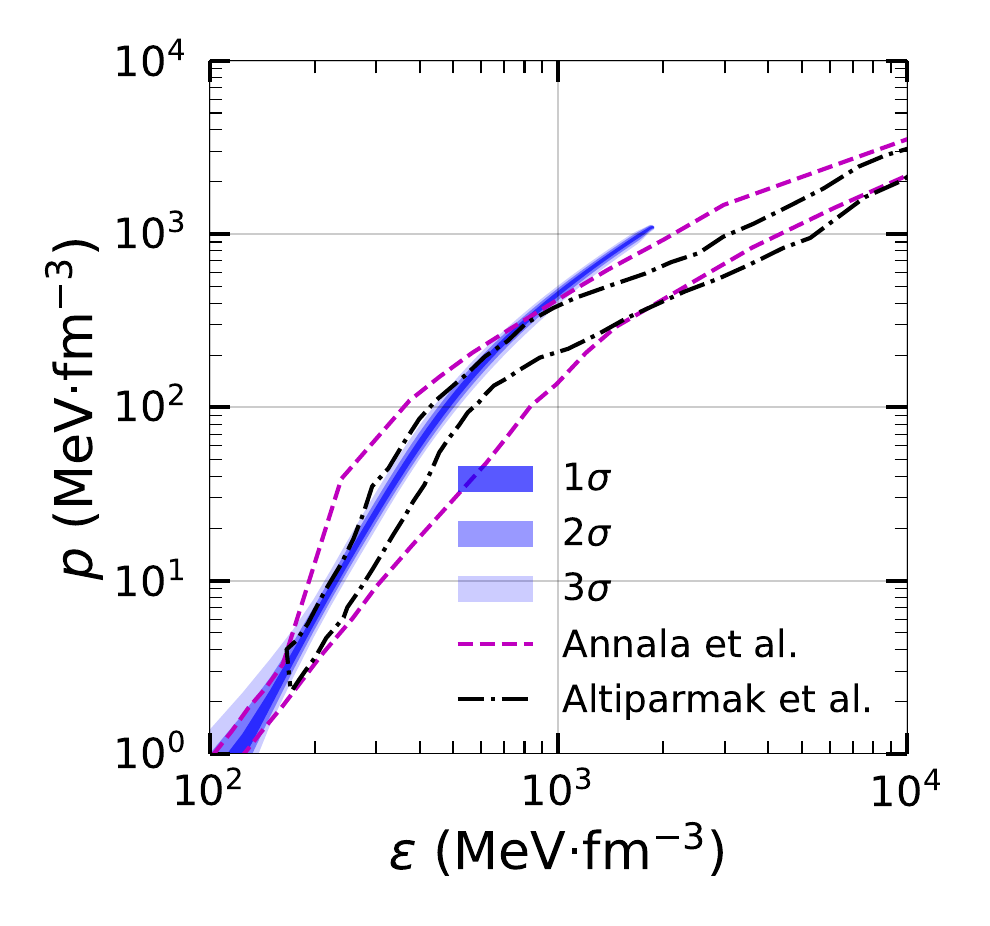}
    \includegraphics[width=0.45\linewidth]{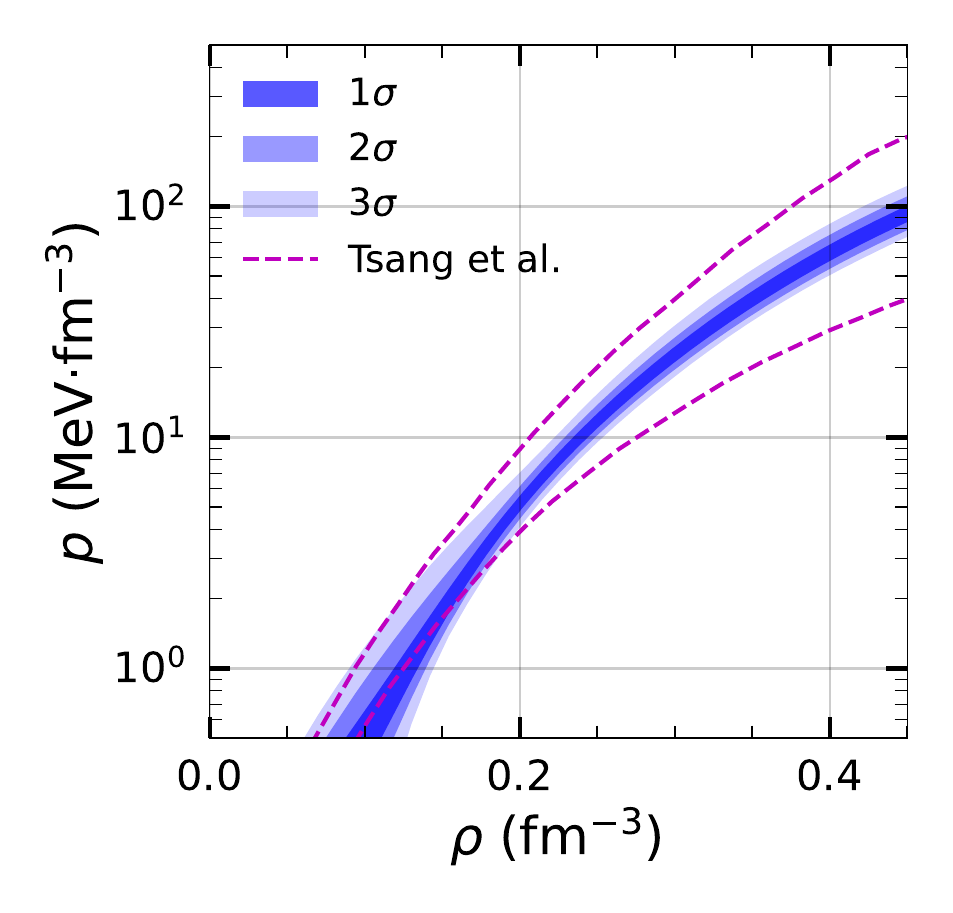}
    \caption{The variation of pressure with respect to energy density (left) and baryon density (right). The different shades of blues show the $1\sigma$, $2\sigma$ and $3\sigma$ \ac{ci}s. In the left panel, the red dashed curve shows the accepted domain for the \ac{eos}s as described in the Ref. \cite{Annala:2021gom} and the black dot-dashed curve is showing the more constrained domain for the \ac{eos}s as studied in the Ref. \cite{Altiparmak:2022bke}. In the right panel, the upper and lower magenta dashed curves are the limits for the pressure at various densities as discussed in the Ref. \cite{Tsang:2023vhh}.}
    \label{fig:eosset_and_eosset_anbpsr}
\end{figure*}

\subsection{\ac{eos} of nuclear matter} \label{sec:eos_of_nuclear_matter}
Here, in this section, we discuss the \ac{eos} set, which is the core focus of the present work. In the previous section, we discussed the coupling parameters evaluated from the Bayesian approach, which is widely used in \ac{ns} physics. In FIG. \ref{fig:eosset_and_eosset_anbpsr}, we plot the 90\% \ac{ci} of the variation of pressure, Eq. (\ref{pressure.nm}), as a function of energy density, Eqs. (\ref{energy.density.nm}), (in the left panel) and baryon density, Eq. (\ref{baryon.density}) (in the right panel) as the blue shaded regions corresponding to the coupling parameter set. In the left panel, the solid magenta and dot-dashed blue curves define the constrained regions for the pressure at a given energy density obtained in Refs. \cite{Annala:2021gom, Altiparmak:2022bke} respectively. The present \ac{eos} set is satisfying the previous results nicely. Similarly, in the left panel, the dashed curves are obtained in the Ref. \cite{Tsang:2023vhh} for pressure as a function of baryon density. We see that the present \ac{eos} set satisfies these results too. Here, we are plotting only the dense matter \ac{eos}. For the case of \ac{ns}, we join this core \ac{eos} with the crust \ac{eos}, which we will discuss later in more detail. Next, we discuss the corresponding speed of sound, $c_s^2 = \frac{dp}{d\epsilon}$.
\begin{figure}
    \centering
    \includegraphics[width=0.95\linewidth]{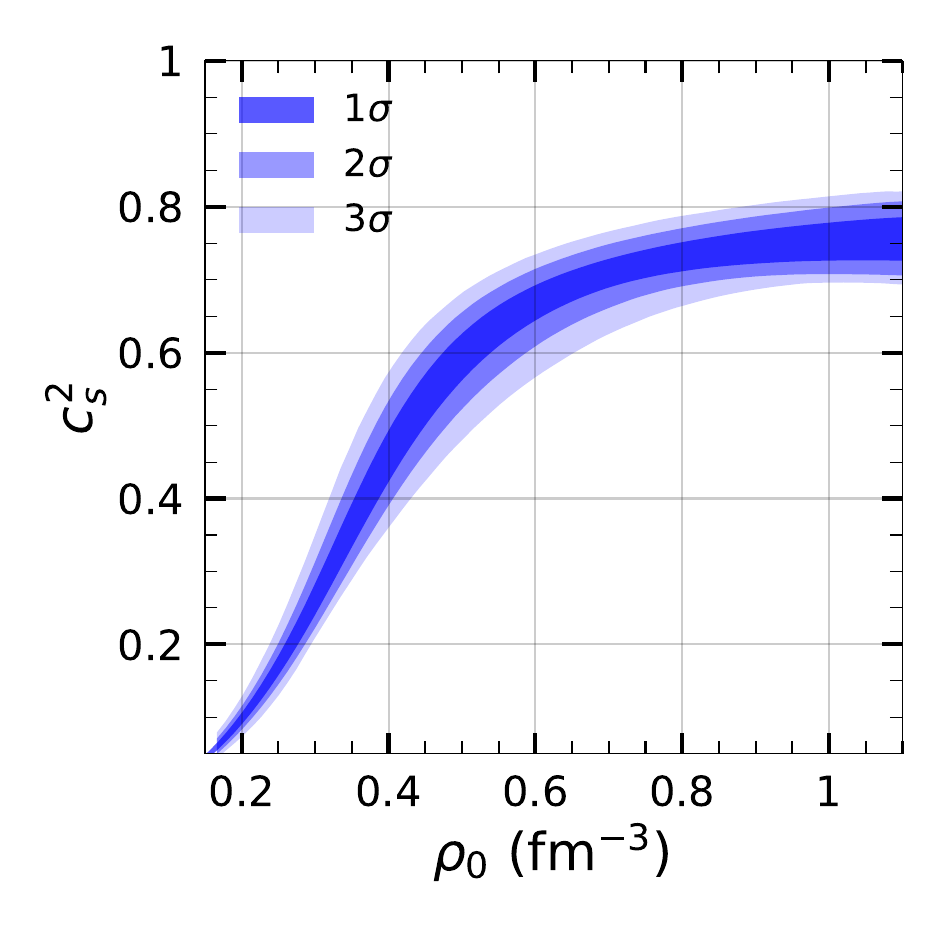}
    \caption{The variation of square of speed of sound as a function of baryon density. Different shades of blues show the $1\sigma$, $2\sigma$ and $3\sigma$ \ac{ci}s.}
    \label{fig:cs2set}
\end{figure}

In the \ac{rmf} model, the speed of sound ($c_s^2 = dp/d\epsilon$), which reflects how pressure responds to changes in energy density, plays a crucial role in determining the stiffness of the \ac{eos} and, consequently, the structure of \ac{ns}s. In the present study, $c_s^2 \le 1$ is also taken as a constraint. The parameters, which give super-luminal \ac{eos} i.e. $c_s^2 \geq 1$ are rejected. In FIG. \ref{fig:cs2set} we plot the 90\% \ac{ci} of the variation of speed of sound ($c_s^2 = dp/d\epsilon$) as a function of baryon number density. We see that the speed of sound exhibits a rapid increase in $c_s^2$ at small densities, while it rises but at a slower pace in intermediate densities, which shows that the nuclear matter becomes extremely dense. We also see that the curves reach a maximum value of $c_s^2$ at high densities. The behaviour of $c_s^2$ is strongly influenced by the meson-nucleon couplings, particularly $g_{\sigma}$ and $g_{\omega}$ interactions, and their nonlinear self-couplings. A stiffer \ac{eos}, often associated with stronger vector couplings, results in a higher speed of sound, enabling the support of more massive \ac{ns}s.

\begin{figure*}
    \centering
    \includegraphics[width=0.45\linewidth]{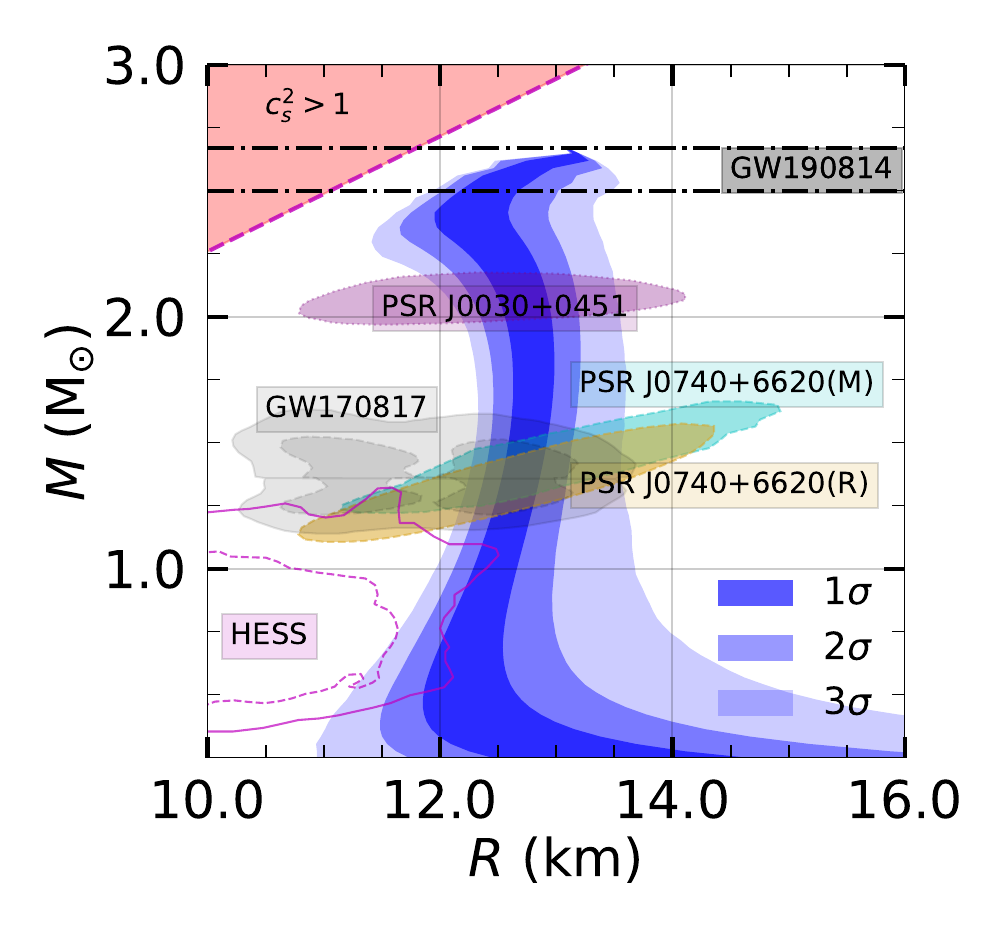}
    \includegraphics[width=0.45\linewidth]{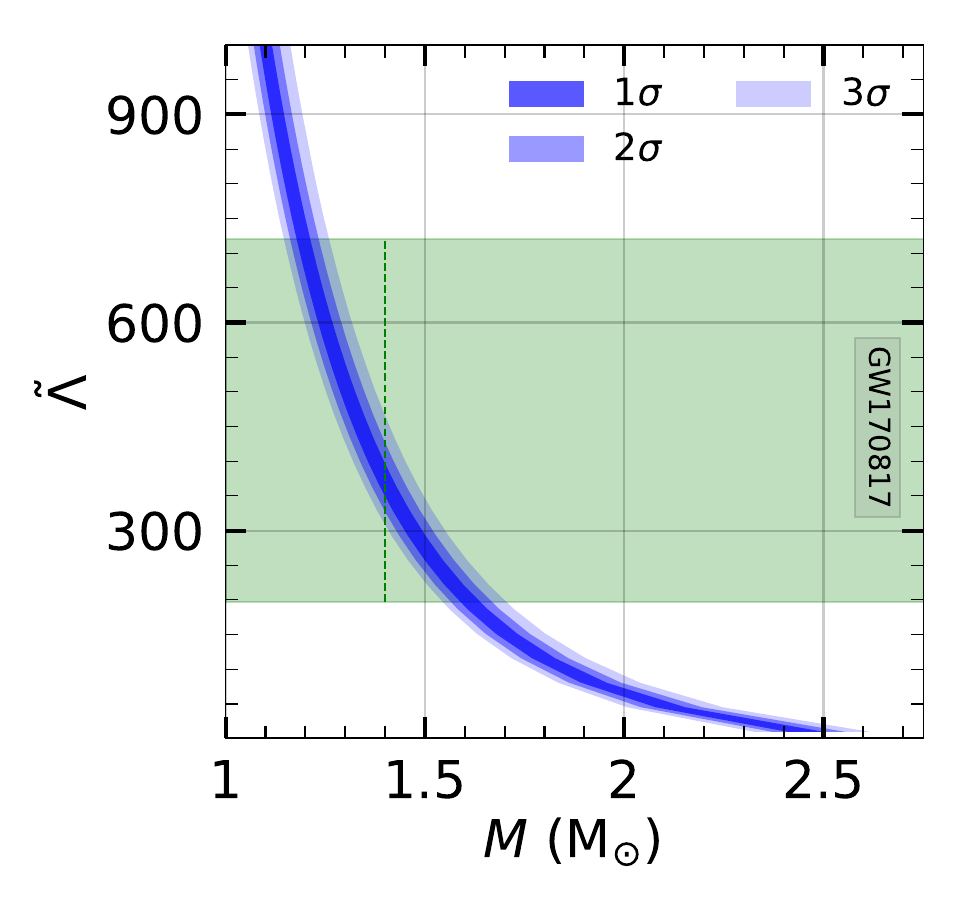}
    \caption{The mass-radius relations in the left panel and weighted tidal deformability as a function of \ac{ns} mass in the right panel. The different blue shaded regions show the $1\sigma$, $2\sigma$ and $3\sigma$ \ac{ci}s. In the left panel, the gray shaded patch corresponds to the GW170817 observation while the magenta curves shows the HESS findings. The magenta shaded region represents the NICER observation of a high mass pulsar PSR J0030+0451 \cite{Miller:2019cac, Riley:2019yda} while the yellow and cyan regions correspond to the NICER observation of the same pulsar PSR J0740+6620 but by different groups \cite{Miller:2021qha, Dittmann:2024mbo}. Two horizontal dot-dashed black lines represent the recent \ac{gw} observation, GW190814. The red shaded region in the plot shows the causality where $c_s^2 \geq 1$. In the left panel, the green shaded region shows the \ac{gw} observation, GW170817 and a vertical dashed green line is drawn at $M=1.4\ {\rm M}_{\odot}$.}
    \label{fig:amrset}
\end{figure*}

\subsection{\ac{ns} properties} \label{sec:ns_properties}
Here, in this section, we discuss the \ac{ns} properties e.g. mass, radius and tidal deformability. For each coupling parameter set, we obtain the \ac{eos} at high densities relevant for the core of \ac{ns}s. For the crust part of the \ac{eos} in each parameter we have taken the BPS \ac{eos} \cite{Baym:1971pw} upto $\rho = 0.04\ {\rm fm}^{-3}$ for the outer crust and then after we use an parameterized \ac{eos} i.e. $p(\rho) = a + b \rho^{\gamma}$ for the inner crust, where $\gamma=4/3$ \cite{Fortin:2016hny}. The values of $a$ and $b$ are evaluated by joining the two ends of the outer core and inner crust. Once we find a \ac{eos} for all densities, we can solve the \ac{tov} equations, Eqs. (\ref{tov.mass}, \ref{tov.pressure}) to obtain the mass-radius relationships and Eqs. (\ref{tidal.ay}, \ref{love_number}, \ref{tidal_weighted}) to obtain weighted tidal deformability. 

In FIG. \ref{fig:amrset}(left), we plot 90\% \ac{ci} of mass-radius curves for all the parameter sets obtained in this study as a blue shaded region. The light and dark grey patch with dotted outline denotes the GW170817 observations \cite{LIGOScientific:2017vwq}. The yellow with dashed outlines denotes the observations of PSR J0030-0470 by \ac{nicer} \cite{Riley:2019yda} while the cyan with dashed outlines shaded regions denote the \ac{nicer} observation of the same pulsar by another group \cite{Miller:2019cac}. The solid magenta with dotted outline region shows the observation of the pulsar PSR 0740+6620 \cite{Miller:2021qha}. Two dot-dashed blue horizontal lines show a recent GW190419 observation \cite{LIGOScientific:2020zkf}  which is still debatable. The straight dashed magenta line is, corresponds to the \ac{eos} $p=\epsilon$ or $c_s^2=1$, showing the limits for the mass-radius, which we call a super-luminal line. If the corresponding mass-radius curve crosses this super-luminal line, the \ac{eos} set is rejected here. From this figure, we see that the parameter set, obtained here, is consistent with the \ac{gw} and \ac{nicer} observations and is below the super-luminal line. In FIG. \ref{fig:amrset} (left), we plot 90\% \ac{ci} of the variation of the dimensionless tidal deformability $\tilde{\Lambda}$ as a function of mass. We see that this result is also consistent with the GW170817 observations \cite{LIGOScientific:2017vwq}.

\section{Summary and Conclusions} \label{sec:summary_and_conclusion}
In the \ac{ns} physics, the \ac{eos} of dense matter is the most important quantity. Because of various uncertainties at higher densities, a few times the nuclear saturation density ($\rho_0$), the \ac{eos} of dense matter is still a big question. In this article, we constrain the nuclear matter \ac{eos} using the Bayesian approach such that the \ac{eos} must satisfy nuclear saturation properties, \ac{ns} experimental observations and while keeping in mind that the \ac{eos} must be sub-luminal. 

We study nuclear matter within a \ac{rmf} theory of nucleons where the interactions are governed by exchange of various mesons e.g. $\sigma$, $\omega$ and $\rho$ with non-linear in $\sigma$, self-interaction of $\omega$ and cross-interaction between $\omega$ and $\rho$ terms with $\Lambda_{\omega\rho}$ coupling parameter at a fundamental level. Here, we consider that the couplings are constant over all the densities considered. We find several models, \href{https://u.pcloud.link/publink/show?code=kZPL3m5Zc5a0w45VVl8Jhk2ragtDFh3l2JM7}{EOS DATA SET}, constrained with various nuclear saturation properties and \ac{ns}s astrophysical observations as listed in TABLE \ref{table-nmsp}. All sets satisfy the nuclear saturation properties and \ac{ns} experimental observations e.g. \ac{nicer}, \ac{gw} and HESS observations. In the present study, we have seen that the cross-coupling parameter $\Lambda_{\omega\rho}$ plays an important role in obtaining a higher mass \ac{ns}. The parameter  $\Lambda_{\omega\rho}$ becomes to stiffens the \ac{eos} at higher densities while the non-zero value of it does not affect much the nuclear saturation properties at saturation densities. Another important coupling parameter is $\lambda$, which takes negative value to produce the high mass \ac{ns}s. In this study, we obtained few sets of parameters which give a sufficiently stiffen \ac{eos}s at high densities and soft at low densities in such a way that the corresponding \ac{eos}s could produce the \ac{ns}s of masses more than $2.5\ {\rm M}_{\odot}$ i.e. we can define the recent \ac{gw} observation \cite{LIGOScientific:2020zkf} which is still debatable. These sets of parameters and corresponding \ac{eos}s can be found from the \href{https://u.pcloud.link/publink/show?code=kZPL3m5Zc5a0w45VVl8Jhk2ragtDFh3l2JM7}{EOS DATA SET}.

\section*{Acknowledgments}
We thank Dr. M. Bhuyan for the valuable discussions throughout this work.

\bibliographystyle{unsrt}
\bibliography{zzbib}


\end{document}